\documentclass[aps,preprint,groupedaddress,showpacs]{revtex4}

\newcommand{\ab}{\textit{ab initio }}
\newcommand{\abb}{\textit{Ab initio }}

\newcommand{\La}{LaMnO$_3$ }
\newcommand{\Lap}{LaMnO$_3$}

\newcommand{\AxBy}{A$_{1-x}$B$_{x}$MnO$_3$ }

\newcommand{\AxByp}{A$_{1-x}$B$_{x}$MnO$_3$}

\newcommand{\LaCax}{La$_{1-x}$Ca$_x$MnO$_3$ }

\newcommand{\LaCa}{La$_{0.5}$Ca$_{0.5}$MnO$_3$ }

\newcommand{\PrCasix}{Pr$_{0.60}$Ca$_{0.40}$MnO$_3$ }
\newcommand{\PrCasixp}{Pr$_{0.60}$Ca$_{0.40}$MnO$_3$}
\newcommand{\PrSr}{Pr$_{0.5}$Ca$_{0.5}$MnO$_3$ }

\newcommand{\LaCap}{La$_{0.5}$Ca$_{0.5}$MnO$_3$}

\newcommand{\eg}{e$_g$ }

\newcommand{\ooneionp}{O$^{-}$}
\newcommand{\oion}{O$^{2-}$ }

\newcommand{\jtwo}{J$_{2}$ }
\newcommand{\jthree}{J$_{3}$ }
\newcommand{\jonep}{J$_{1}$}
\newcommand{\jtwop}{J$_{2}$}

\newcommand{\jonemag}{$|$J$_{1}|$ }

\newcommand{\dt}{d$^{3}$ }
\newcommand{\df}{d$^{4}$ }
\newcommand{\dfp}{d$^{4}$}

\newcommand{\CEx}{C$_{x}$E$_{1-x}$ }

\usepackage{graphicx}
\usepackage{latexsym}

\begin{document}

\title{ Spin Polaron Effective Magnetic Model for \LaCa }

\author{N.P. Konstantinidis}

\affiliation{ Department of Physics and Department of Mathematics, \\
University of Dublin, Trinity College, Dublin 2, Ireland}

\author{C.H. Patterson}

\affiliation{ Department of Physics and Centre for Scientific Computation,\\
University of Dublin, Trinity College, Dublin 2, Ireland}

\date{\today}

\begin{abstract}
The conventional paradigm of charge order for \LaCax for $x=0.5$ has been
challenged recently by a Zener polaron picture emerging from experiments and
theoretical calculations. The effective low energy Hamiltonian for the magnetic
degrees of freedom has been found to be a cubic Heisenberg model, with ferromagnetic nearest neighbor and frustrating antiferromagnetic next nearest neighbor interactions in the planes, and antiferromagnetic interaction between planes. With linear spin wave theory and diagonalization of small clusters up to $27$ sites we find that the behavior of the model interpolates between the A and CE-type magnetic structures when a frustrating intraplanar
interaction is tuned. The values of the interactions calculated by \ab methods indicate a possible non-bipartite picture of polaron ordering differing from the conventional one.
\end{abstract}


\maketitle

\section{Introduction}
Charge, orbital and spin order in doped manganites (\AxByp, where A is a
trivalent ion and B a divalent ion) with $x \geq$ 0.4 is generally discussed in
terms of the Goodenough model \cite{Goodenough55} and the double exchange
Hamiltonian \cite{Dagotto02}.  In that picture, charge ordering (CO) and
orbital ordering (OO) is associated with 3d populations of Mn ions, which may
be \dt or \dfp. In \LaCap, for example, it is believed that there are equal
numbers of Mn ions with \dt and \df populations in the CO state
\cite{Radaelli97}. However, recent experimental and theoretical investigations point to a more complex picture, and there is evidence that a Zener polaron or similar model, which is quite distinct from the original Goodenough model, describes the CO state of \LaCa and \PrCasixp \cite{Daoud02,Zheng03,Efremov03}. Here a frustrated Heisenberg Hamiltonian is used to model the Zener polaron state of half-doped manganites using spin wave theory and exact diagonalization of clusters with periodic boundary conditions.  The parameter space for the Hamiltonian is derived from \ab calculations on the Zener polaron state of \LaCa \cite{Zheng03}. In particular, we report energetics, spin waves, magnetization, spin correlation functions, the spin contribution to heat capacity and magnetic susceptibility for this model.

In the Zener polaron picture pairs of Mn ions are tightly bound into ferromagnetic (FM) dimers which interact relatively weakly with each other \cite{Daoud02,Zheng03}. The Mn ion 3d population is \df on all Mn ions and electrons are transferred from \oion ions located between Mn ions in dimers, to form Mn-\ooneionp-Mn polarons. Electron transfer is necessary to satisfy electron counting. The dimers (Zener polarons) have spin 7/2. Their electronic structure is reflected in the $\it{Pm}$ crystal structure of \PrCasix determined by neutron scattering from a single crystal \cite{Daoud02}.  The $\it{P2_{1}/m}$ structure for \LaCa determined by neutron scattering \cite{Radaelli97} from a powder is different in several respects. The crystal structure of \LaCa has been calculated by using \ab methods to minimize the crystal total energy \cite{Patterson03} and the resulting structure resembles that for \PrCasix \cite{Daoud02}. The Zener polaron picture for half-doped manganites may therefo
re apply more generally \cite{Rivadulla02}.
In the frustrated Heisenberg Hamiltonian used in this work, polarons are treated as single magnetic units with spin 7/2. This is appropriate for Mn ions tightly bound into FM dimers when the internal FM coupling greatly exceeds the inter-polaron coupling. \abb calculations on \LaCa show that magnetic coupling within polarons is strong and ferromagnetic (J $\sim 200$ meV), while coupling between polarons along their zig-zag chains is also FM but much weaker (J $\sim 10$ meV) \cite{Zheng03}. There are both antiferromagnetic (AF) and FM couplings of the same order of magnitude between polarons in neighboring chains so that interchain coupling is strongly frustrated. 

\begin{figure}
\includegraphics[width=5in,height=5in]{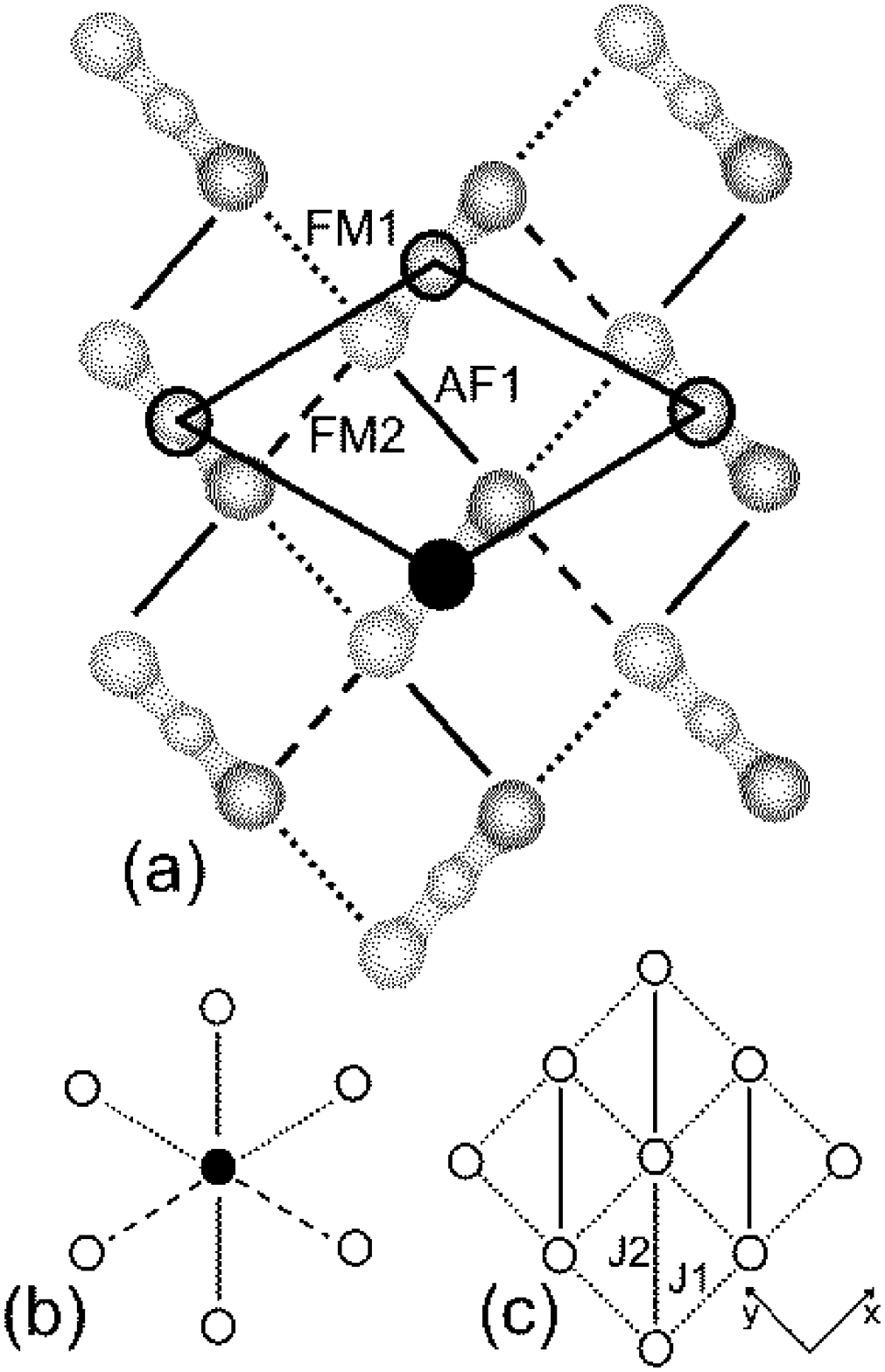}
\caption{Schematic illustration of the connectivity of the model: (a) effective
exchange interactions between polarons (FM stands for ferromagnetic and AF for
antiferromagnetic). (b) Magnetic unit cell. (c) Connectivity of the model in
the planes. Interplanar interactions are AF.}
\vspace{20pt}
\label{fig:1}
\end{figure}

Zener polarons and exchange couplings in \LaCa are shown schematically in
Fig. \ref{fig:1}, together with the magnetic unit cell. Intrachain FM
coupling is labelled FM1.  Competing AF and FM intraplanar, interchain couplings are labelled AF1 and FM2; there is also an AF interplanar coupling, AF2, which is not shown. \abb calculations on \La show that exchange coupling constants in that compound depend on OO and that not only the magnitudes, but also the signs of exchange constants can change when OO changes \cite{Nicastro02}. In both \La and \LaCap, AF couplings are found between neighboring Mn ions when their filled \eg orbitals are both oriented perpendicular to a Mn-O-Mn axis, while FM couplings are found between Mn ions when their filled \eg orbitals are oriented
perpendicular to each other, one being oriented along the Mn-O-Mn axis and the
other perpendicular to it. FM exchange constants FM1 and FM2 in \LaCa were found to be similar in magnitude by \ab calculation ($-12$ and $-14$ meV, respectively) \cite{Zheng03}; they are assumed to have the same value, \jonep, in this work. AF1 and AF2 exchange constants were found to be 5 and 8 meV, respectively, and correspond to \jtwo and \jthree in the model described below. It is noted that Unrestricted Hartree-Fock (UHF) calculations used to obtain these exchange constants underestimate the magnitude of the AF exchange constants in \Lap, while the FM exchange constant is in agreement with the experimental value \cite{Nicastro02}. Inspection of Fig. \ref{fig:1}(a) shows that each polaron is ferromagnetically coupled to four neighbors, two on each side, and antiferromagnetically coupled to two neighbors in the same plane (Fig. \ref{fig:1}(b)), along the diagonal direction; this arrangement of exchange couplings is equivalent to the model shown in Fig. \ref{fig:1}(c).

Half-doped manganites possess either A, CE or \CEx type AF
magnetic states below their N\'eel temperatures, depending on the identities of
the ions A and B in \AxBy \cite{Kajimoto02}. CE-type AF order consists of
zig-zag FM chains antiferromagnetically coupled to neighboring chains. \CEx order is an incommensurate charge and orbital CE-type order \cite{Wollan55}. CE and \CEx order is found for wider gap manganites such as \PrCasix \cite{Daoud02} and \LaCa \cite{Radaelli97}, while A-type order is found for metallic manganites such as \PrSr \cite{Kajimoto02}. Incommensurability and a fine energetic balance between A and CE-type magnetic order are therefore features of the half-doped manganites which also appear in \ab calculations \cite{Zheng03}. Magnetic susceptibility data for \PrCasix \cite{Daoud02} has been interpreted in terms of ordering of magnetic moments of Zener polarons in a CE-type state below T$_N$ = 115 K, and ordering of magnetic moments into Zener polarons at the CO temperature, T$_{CO}$ = 330 K. 

The ground state of the classical, magnetic Hamiltonian corresponding to Fig. \ref{fig:1}(c) is A-type when the magnitude of \jonep is less than half that of \jtwop. The ground state changes to an incommensurate spin spiral state with the in-plane component of the wave vector parallel to the diagonal direction of $J_{2}$ when the ratio of exchange constants \jtwop / \jonemag exceeds 0.5. The spin spiral state becomes the commensurate, orthogonal phase described by Efremov et al. \cite{Efremov03} when the spiral wavevector becomes ($\pi / 2$, $\pi / 2$, $\pi$); the classical magnetic model predicts that this state is found only in the limit \jtwop$/$\jonemag $\to$ $\infty$. Both spin wave theory and cluster diagonalization support a picture in which
the magnetic structure in half-doped manganites is strongly dependent on the
relative magnitudes of exchange constants $J_{1}$ and $J_{2}$. Specifically, AF
coupling between polarons tunes the magnetic correlations between a state
with in-plane ferromagnetism, similar to the A-type structure, and a state
where AF correlations are dominant. The latter state is the spin
spiral, which is reminiscent of the CE-type structure. A similar picture has
emerged in \cite{Efremov03} with non-bipartite magnetic structures competing
for the ground state. These authors find that the magnetic structure of the
ground state around half-filling is subtly dependent on the extent of doping,
$x$, and for a part of the phase diagram the ground state is intermediate
between a conventional CO state and a Zener polaron state.

A two-dimensional Heisenberg model, with frustrated, AF interactions similar to those in the three-dimensional model used here, describes the magnetic properties of organic molecular crystals \cite{Trumper99,Zheng99}. A Heisenberg model with the same connectivity as that used here has been applied to $\alpha^{'}$-NaV$_{2}$O$_{5}$ \cite{Suaud00}. A similar spin wave approach to the one used in this paper was recently applied to \LaCa \cite{Ventura03}.  In that work $\it{each}$ Mn ion was assigned a spin $3/2$ or $2$, there were FM couplings within zig-zag chains and AF couplings between spins in adjacent chains.  Hence that work differs from the present work in the magnetic units used (single ions versus dimers) and the presence or absence of frustration in interchain coupling. 

The plan of the rest of this paper is as follows: in section \ref{sec:2} the
model is introduced and solved at the classical level; in section
\ref{sec:3} linear spin wave theory is applied; in section \ref{sec:4} the
results of diagonalization of the Hamiltonian for small clusters with periodic
boundary conditions are presented. Finally conclusions are presented in section
\ref{sec:5}.

\section{Hamiltonian and Classical Solution}
\label{sec:2}
The Heisenberg Hamiltonian used in this work is
\begin{equation}
H = \sum_{i,j,k} [ \textrm{ } J_{1} \textrm{ } ( \vec{s}_{i,j,k} \cdot
\vec{s}_{i+1,j,k} + \vec{s}_{i,j,k} \cdot \vec{s}_{i,j+1,k} ) + J_{2}
\textrm{ } \vec{s}_{i,j,k} \cdot \vec{s}_{i+1,j+1,k} + J_{3} \textrm{ }
\vec{s}_{i,j,k} \cdot \vec{s}_{i,j,k+1} \textrm{ } ] \textrm{ }.
\label{eqn:1}
\end{equation}
The FM zig-zag chains are defined by consecutive steps in the $\hat{x}$ and
$-\hat{y}$ directions in Fig. \ref{fig:1}(c). $z$ is the interplanar axis.
$i,j$ and $k$ label sites along the three axes, $J_{1}$ is FM
and negative while $J_{2}$ and $J_{3}$ are AF and positive. Interactions within
$xy$ planes are frustrated since $J_{1}$ prefers parallel spin alignment while
$J_{2}$ favors antiparallel alignment along the diagonals.
The term bond is used to describe these interactions from here on. Ratios of the magnitudes of \ab exchange constants are $J_{2}/|J_{1}| = 0.38$ and $J_{3}/|J_{1}|=0.62$ (using an average value of the FM1 and FM2 of $-13$ meV for \jonep) \cite{Zheng03}. The interplanar exchange coupling \jthree is not frustrated and the ratio $J_{3}/|J_{1}|$ will be fixed at 0.5 from now on, except where noted.

At the classical level \cite{Villain59,Yoshimori59}, the angle between
neighboring spins is found by minimizing the structure factor,
\begin{equation}
J(\vec{q}) = J_{1} \textrm{ } ( cosq_{x} + cosq_{y} ) + J_{2} \textrm{ }
cos(q_{x}+q_{y}) + J_{3} \textrm{ } cosq_{z} .
\label{eqn:2}
\end{equation}
The solution is $\vec{q}=(0,0,\pi)$ for $J_{2}/|J_{1}| \leq $ 0.5 and
$\vec{q}=(q,q,\pi)$ with $q=$arccos$(-J_{1}/2J_{2})$ when
$J_{2}/|J_{1}| > $ 0.5. The former case corresponds to FM intraplanar
order and AF interplanar order; the latter corresponds to a spin spiral
in the plane, with the intraplanar component of the wavevector along the direction of the $J_{2}$ bond. The angle between neighboring spins is $q$.  In the limit $J_{2}/|J_{1}| \to \infty$, $q \to \pi/2$ so that neighboring spins are at right angles; this is the orthogonal phase described in \cite{Efremov03}.
Thus the solution to Eq. \ref{eqn:1} for small $J_{2}/|J_{1}|$ corresponds to A-type AF order, while for $J_{2}/|J_{1}| \to \infty$ the structure is a noncolinear CE-type magnetic phase.

Clusters of cubic symmetry with $2 \times 2 \times 2$ and $3 \times 3 \times 3$ sites will be considered in exact diagonalization calculations in section \ref{sec:4}. For the second cluster, periodic boundary conditions are frustrated for the AF bonds, due to the odd number of spins along the corresponding directions. For both clusters the classical ground state is FM in the $xy$ plane for $J_{2}/|J_{1}| \leq 1$, with energies per bond $J_{1}$ and $J_{2}$. For $J_{2}/|J_{1}| > 1$, the bond energies are $0$ and $-J_{2}$ for the $8$-site cluster, and $J_{1}/4$ and $-J_{2}/2$ for the $27$-site cluster. For the $8$-site cluster this is orthogonal type of magnetic order \cite{Efremov03}, while for the $27$-site cluster it is a frustrated configuration along the diagonals. Along the $z$ direction the energy per bond is $-J_{3}$ for $8$ sites and $-J_{3}/2$ for $27$, regardless of the value of $J_{2}/|J_{1}|$. Frustration in boundary conditions increases the energy for the $27$-site cluster. The total energies are $4 ( 2 J_{1} + J_{2} - J_{3} )$ for $8$ sites and $27 ( 2 J_{1} + J_{2} - (J_{3}/2) )$ for $27$ sites when $J_{2}/|J_{1}| \leq 1$, and the corresponding energies for $J_{2}/|J_{1}| > 1$ are $-4(J_{2}+J_{3})$ and $(27/2)(J_{1}-J_{2}-J_{3})$ respectively. The ratio $J_{2}/|J_{1}|$ where the ground state correlations change character is different compared with the infinite lattice value due to the finite size of the clusters.

\section{Linear Spin Wave Theory}
\label{sec:3}
\begin{figure}
\includegraphics[width=5in,height=5in]{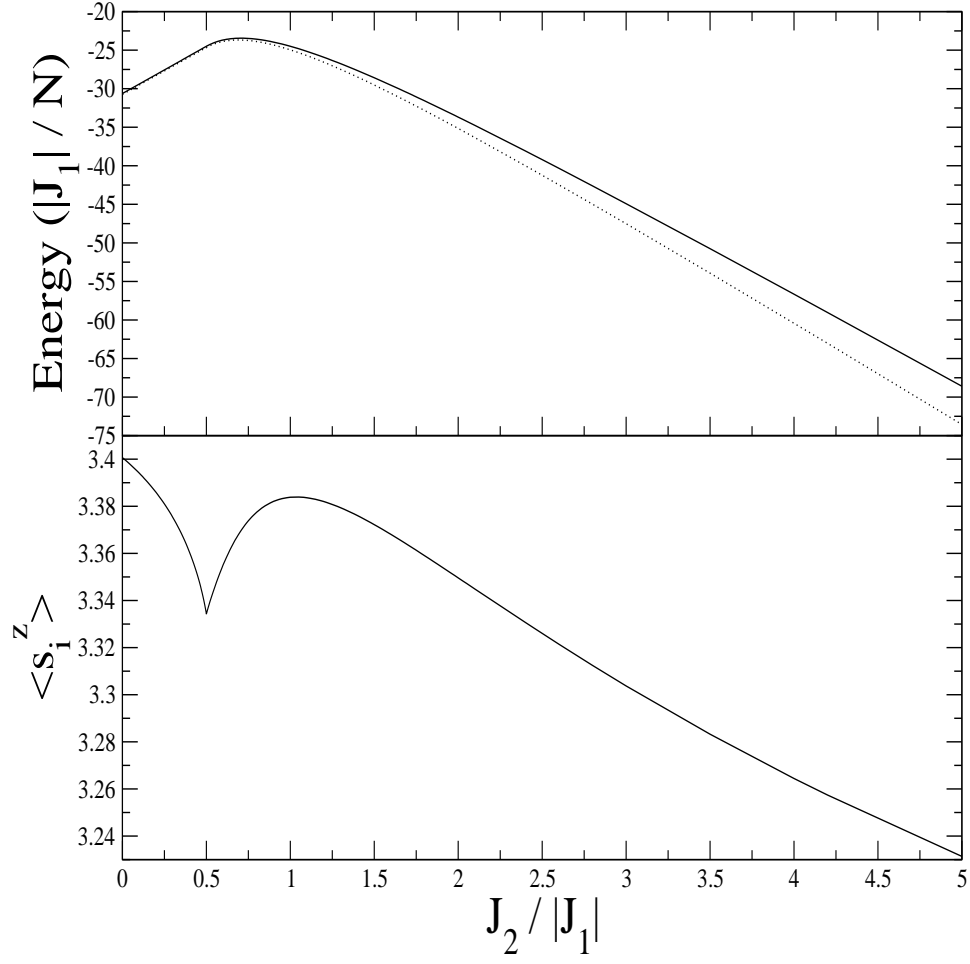}
\caption{Ground state energy and magnetization per site as a function of
$J_{2}/|J_{1}|$ for $J_{3}/|J_{1}|=0.5$ within linear spin wave theory. Top:
solid line: classical energy, dotted line: spin wave energy.}
\label{fig:2}
\end{figure}

Owing to the spiral nature of the classical ground state for
$J_{2}/|J_{1}| > $ 0.5, it is convenient to redefine the $s_{i}$
operators so that the local quantization axis points along the
the classical solution spin directions. Only one type of bosonic operator is necessary here,
although the Hamiltonian becomes more complex. After introducing the
Holstein-Primakoff transformation \cite{Holstein40} with operators $a$ and
Fourier transforming, the Hamiltonian becomes
\begin{equation}
H = N E_{cl} + S \sum_{\vec{k}} [ A_{\vec{k}} a_{\vec{k}}^{\dag} a_{\vec{k}} +
\frac{B_{\vec{k}}}{2} ( a_{\vec{k}}^{\dag} a_{-\vec{k}}^{\dag} + a_{\vec{k}}
a_{-\vec{k}}) ],
\label{eqn:3}
\end{equation}
where $N$ is the number of sites, $E_{cl} = S^{2} \displaystyle \sum_{i} J_{i}
cos\theta_{i}$, $A_{\vec{k}} = - \displaystyle \sum_{i} J_{i} [ 2 \textrm { }
cos\theta_{i} - ( 1 + cos\theta_{i} ) \textrm{ } cos(\vec{k} \cdot
\vec{\delta_{i}}) ]$ and $B_{\vec{k}} = - \displaystyle \sum_{i} J_{i} ( 1 -
cos\theta_{i} ) \textrm{ } cos(\vec{k} \cdot \vec{\delta_{i}})$. $i$ refers to
bonds in the unit cell with $\theta_{i}$ the angle at the classical level
between two spins connected by bond $J_{i}$, and $\vec{\delta_{i}}$ the unit
vector in the bond's direction. After a Bogoliubov transformation to new
operators $\alpha_{\vec{k}}$ \cite{Bogoliubov47,Bogoliubov58}, the diagonalized
Hamiltonian is
\begin{equation}
H = N E_{cl} + S \textrm{ } ( \textrm{ } \sum_{\vec{k}}
\epsilon_{\vec{k}} \alpha_{\vec{k}}^{\dag} \alpha_{\vec{k}} + \sum_{\vec{k}}
\frac{\epsilon_{\vec{k}}-A_{\vec{k}}}{2} \textrm{ } ),
\label{eqn:4}
\end{equation}
with $\epsilon_{\vec{k}}=\sqrt{A_{\vec{k}}^{2}-B_{\vec{k}}^{2}}$. The average
spin magnitude per site along the classical solution direction is
\begin{equation}
<s_{i}^{z}> \textrm{} = s_{i} + \frac{1}{2} - \frac{1}{2N} \sum_{\vec{k}}
\frac{A_{\vec{k}}}{\epsilon_{\vec{k}}}.
\end{equation}
The ground state energy is found by setting the occupation number
$\alpha_{\vec{k}}^{\dag} \alpha_{\vec{k}}$ equal to zero for every $\vec{k}$.
The ground state energy and magnetization per site are plotted in Fig.
\ref{fig:2} as a function of $J_{2}/|J_{1}|$ with $J_{3}/|J_{1}| = 0.5$.
The correction to the energy for quantum fluctuations for $s_{i}$=7/2 is very small owing to the large spin magnitude; the classical spin structure
survives with only minor changes in the magnetization. The point of maximal
frustration where the energy has a maximum is shifted from $0.71$ for the classical case to $0.69$ in linear spin wave theory. Quantum fluctuations lower the local magnetization from 7/2 per site and there is a minimum in the local magnetic moment when the classical ground state changes from a FM to a spiral state.

\begin{figure}
\includegraphics[width=5in,height=5in]{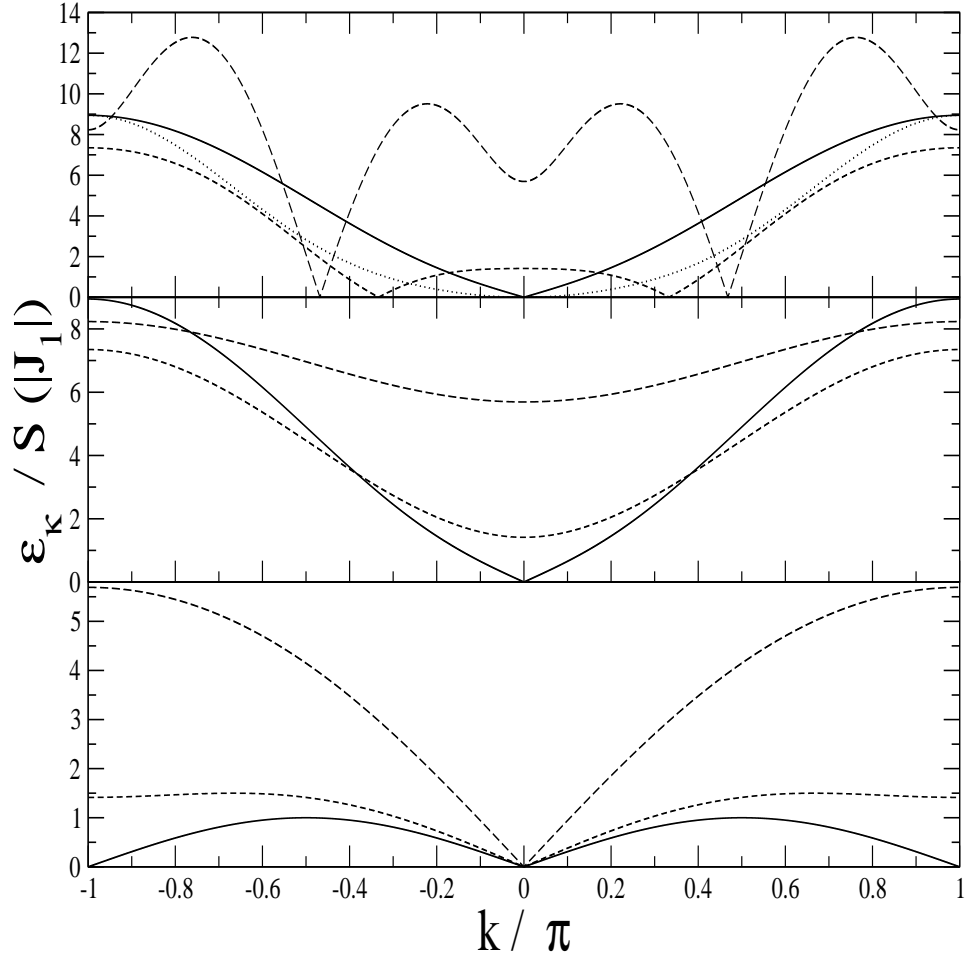}
\caption{Dispersion relations along different directions in the Brillouin zone
for $J_{3}/|J_{1}|=0.5$ within linear spin wave theory. Top: $(k,k,\pi)$,
middle: $(k,-k,\pi)$, bottom: $(0,0,k)$. straight line: $J_{2}/|J_{1}|=0$ for
top and $0 \le J_{2}/|J_{1}| \le 0.5$ for middle and bottom, dotted line:
$J_{2}/|J_{1}|=0.5$ for top, dashed line: $J_{2}/|J_{1}|=1$, long-dashed line:
$J_{2}/|J_{1}|=5$.}
\label{fig:3}
\end{figure}

Spin wave dispersion relations along $(k,k,\pi)$ (parallel to the AF $J_{2}$ bond), $(k,-k,\pi)$ (perpendicular to the $J_{2}$ bond),
and $(0,0,k)$ (perpendicular to planes containing zig-zag chains) are shown
in Fig. \ref{fig:3}. When $J_{2}/|J_{1}| = 0$, dispersion relations are
characteristic of a FM ground state; at the transition point,
$J_{2}/|J_{1}| = 0.5$, the spin wave velocity vanishes along the $(k,k,\pi)$ direction and magnetization
corrections have a local maximum \cite{Merino99}. When $J_{2}/|J_{1}| > 0.5$, changes in the magnetic ground state to a spin spiral become evident in the dispersion relation along $(k,k,\pi)$ (Fig. \ref{fig:3}(a)); zero modes appear at the spiraling wave vector, whose magnitude increases with $J_{2}/|J_{1}|$. The zero mode occurs at $(\pi/2,\pi/2,\pi)$ when $J_{2}/|J_{1}| \to \infty$.
When $J_{2}/|J_{1}| \leq 0.5$, spins are ferromagnetically correlated along the $\hat{x}$ and $\hat{y}$ directions and a dispersion relation characteristic of FM order in the ground state is found (Fig. \ref{fig:3}(b)). But when $J_{2}/|J_{1}| > 0.5$, a gap develops at (0,0,0) since there is a spiral in the planes. AF order from plane to plane along the $(0,0,z)$ direction (Fig. \ref{fig:3}(c)) is reduced due to quantum fluctuations as $J_{2}/|J_{1}|$
increases, with the energy gain increasing along the $J_{2}$ bond.
Inclusion of the interplanar interaction makes the system three dimensional and
the integrals in the calculation of the ground state energy and the magnetization per site are well-behaved, while in the calculation in \cite{Trumper99}
and \cite{Merino99} the planar character of the model enhances the role of
quantum fluctuations \cite{Mermin66}.

\section{Diagonalization of Finite clusters}
\label{sec:4}

Hamiltonians (Eq. \ref{eqn:1}) for three dimensional clusters with periodic
boundary conditions and spin 7/2 or 1/2 were diagonalized after block
factorization using permutation
and spin symmetries \cite{Bernu94,Florek02,Waldmann00}. Hamiltonians for
$2 \times 2 \times 2$ and $3 \times 3 \times 3$ clusters, which have all the
symmetries of the bulk, and a $3 \times 3 \times 2$ cluster were diagonalized.
Energy eigenvalue spectra and nearest and next nearest neighbor correlation
functions were computed. Whenever full diagonalization was not possible, lowest
eigenvalues were calculated using the numerical package ARPACK
\cite{ARPACK,Shaw}. The aims were to corroborate results presented above by
investigating the influence of quantum fluctuations on the classical results,
further study the competition of FM and AF intraplanar and AF interplanar
exchange couplings and examine finite-size effects. Diagonalization of
Hamiltonians with $s_{i}=7/2$ was limited to $8$-site clusters due to computer memory requirements, whereas Hamiltonians for $27$-site clusters could be diagonalized
for $s_{i}=1/2$.

The permutation symmetry group of the Hamiltonian, which is the symmetry group of the clusters in real space, is the product of the translation and the point symmetry groups \cite{Bernu94,Florek02,Waldmann00}. The point symmetry group for the cubic clusters is $D_{2h}$. The Hamiltonian Eq. (\ref{eqn:1}) is also symmetric with respect to time reversal. The spin inversion symmetry group consists of the identity and the spin inversion operation, and its product with the permutation group gives the total symmetry group of the lattice. Eigenstates are characterized by the momentum $\vec{k}$ and the irreducible representation of the total symmetry group that they belong to.

\begin{figure}[ht!]
\includegraphics[width=5in,height=5in]{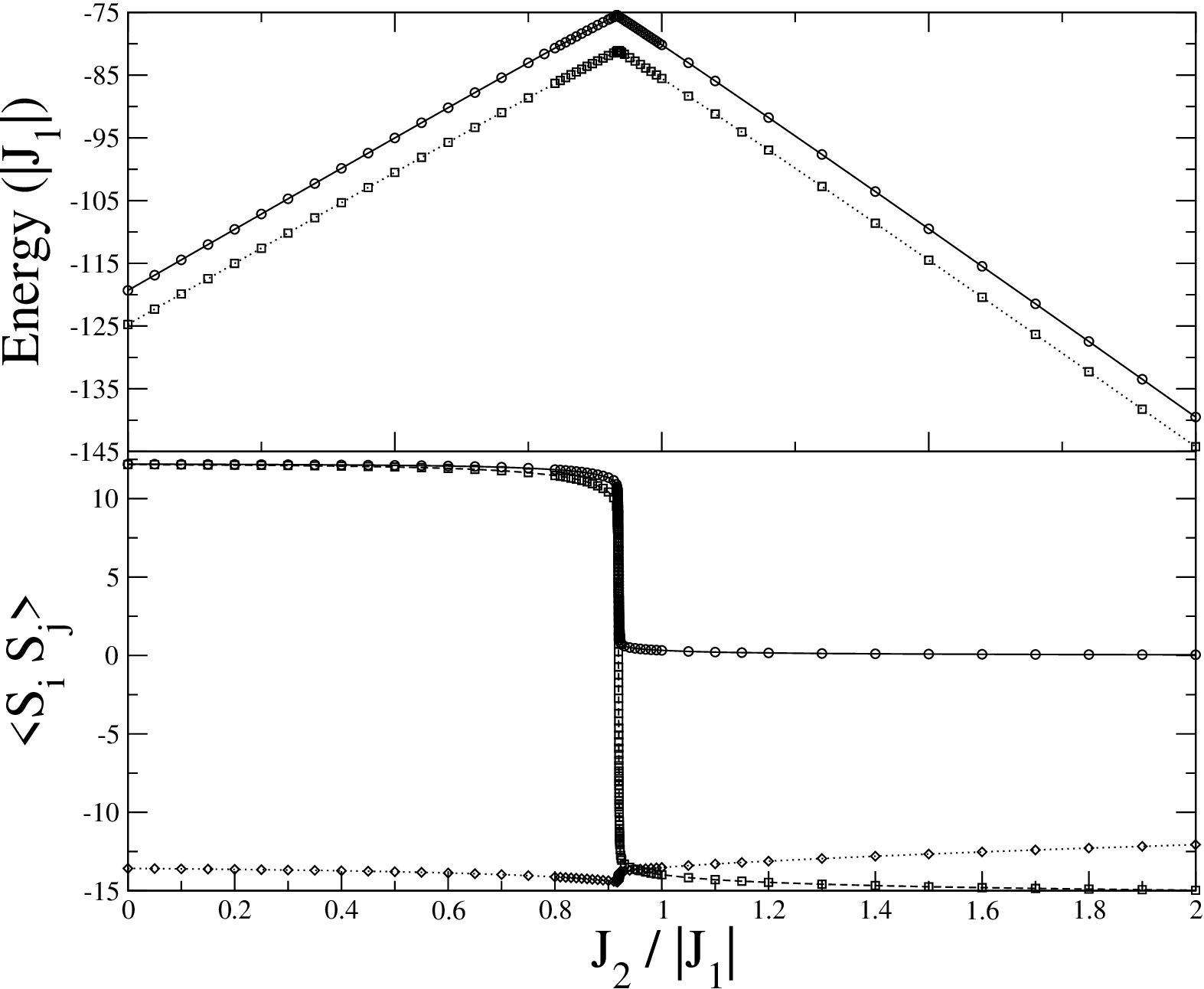}
\caption{Energy and nearest and next nearest neighbor correlation functions of
the ground state for $s_{i}=\frac{7}{2}$ and a $2 \times 2 \times 2$ cluster as
a function of $J_{2}/|J_{1}|$. Top: $J_{3}/|J_{1}|=0.4$ ($\circ$) and
$J_{3}/|J_{1}|=0.5$ ($\Box$). Bottom: correlation function for bond $J_{1}$
($\circ$), for bond $J_{2}$ ($\Box$), and for bond $J_{3}$ ($\diamond$). The
lines are guides for the eye.}
\label{fig:4}
\end{figure}

The ground state energy for the $2 \times 2 \times 2$ $s_{i}=7/2$ cluster is
plotted in Fig. \ref{fig:4}(a) for two values of $J_{3}/|J_{1}|$ as a
function of $J_{2}/|J_{1}|$. It is a total spin $S=0$ state with momentum
$\vec{k}=\vec{0}$ and is symmetric
with respect to spin inversion. Frustration is maximal for $J_{2}/|J_{1}| = 0.91592$ and $0.91895$ when $J_{3}/|J_{1}| = 0.4$ and $0.5$, respectively. The classical transition point $J_{2}/|J_{1}|=1$ is therefore renormalized by quantum fluctuations towards smaller $J_{2}/|J_{1}|$. 

Around the points of maximal frustration the characteristics of the ground
state change, as shown by the correlation functions in Fig. \ref{fig:4}(b). For
small $J_{2}/|J_{1}|$ the nearest neighbor and next nearest neighbor
correlation
functions are strongly FM. For larger $J_{2}/|J_{1}|$ the nearest neighbor correlation function is very small, while the next nearest neighbor correlation function becomes strongly AF. At the same time, the interplanar correlation function, which is always AF, loses some of its strength due to quantum fluctuations, with the spins gaining energy predominantly via the AF $J_{2}$ bond. The changes in the nature of the correlation functions are quite sharp, similar to the behavior of the energy around maximum frustration. The relative changes of the correlation functions as a function of $J_{2}/|J_{1}|$ are very similar to changes in the classical solution. This classical-like behavior is due to the large value of the spin, $s_{i}=7/2$.

\begin{figure}
\includegraphics[width=5in,height=5in]{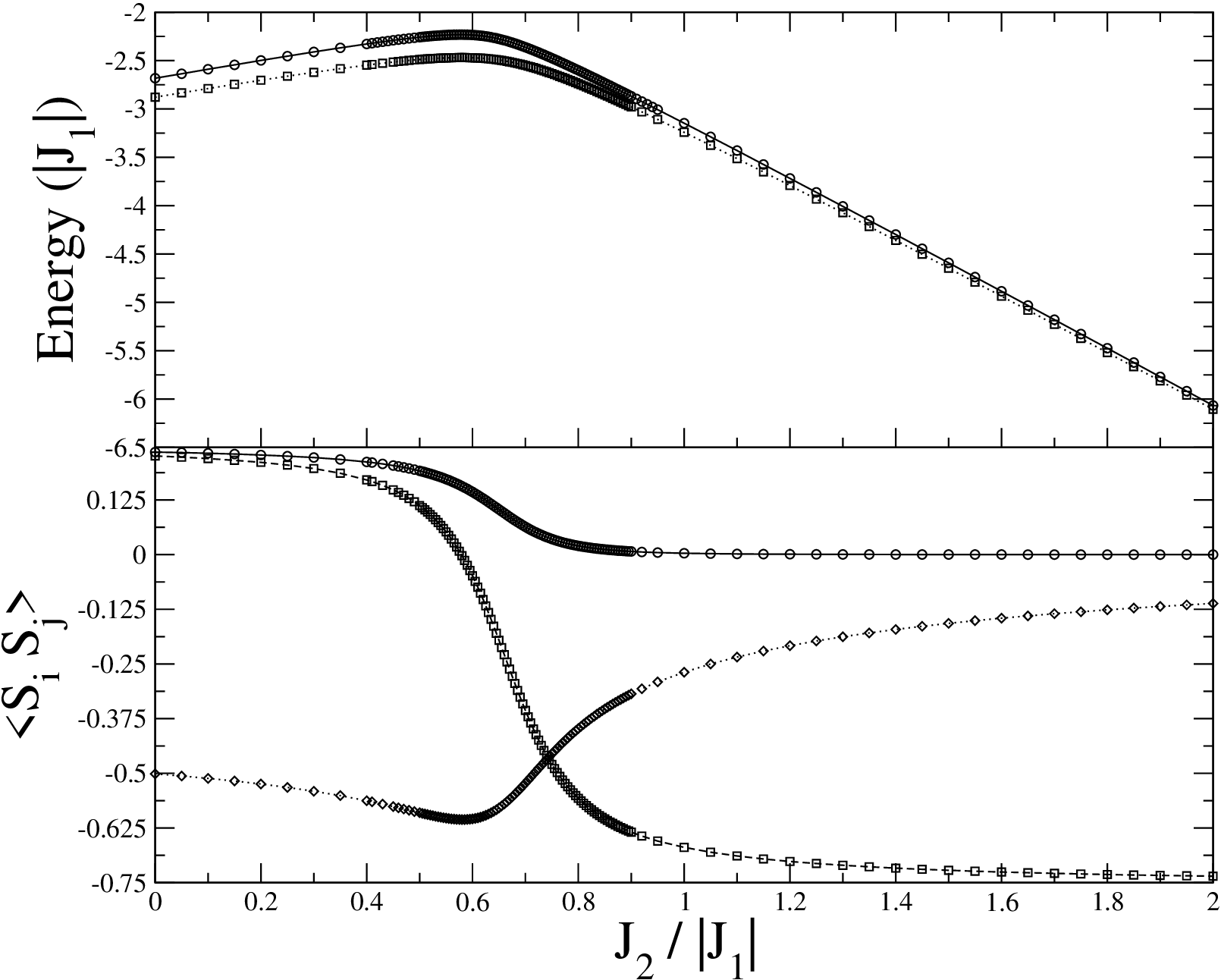}
\caption{Energy and nearest and next nearest neighbor correlation functions of
the ground state for $s_{i}=\frac{1}{2}$ and a $2 \times 2 \times 2$ cluster as
a function of $J_{2}/|J_{1}|$. Top: $J_{3}/|J_{1}|=0.4$ ($\circ$) and
$J_{3}/|J_{1}|=0.5$ ($\Box$). Bottom: correlation function for bond $J_{1}$
($\circ$), for bond $J_{2}$ ($\Box$), and for bond $J_{3}$ ($\diamond$). The
lines are guides for the eye.}
\label{fig:5}
\end{figure}

Calculation of the low energy properties for the $27$-site, $s_{i}=7/2$ cluster is not possible, as noted above. To study the effect of the
frustrating bond $J_{2}$ as a function of cluster size, we consider $s_{i}=1/2$
and estimate the changes compared to the $8$-site case. The ground state energy
for $J_{3}/|J_{1}|=0.4$ and $0.5$ is shown in Fig. \ref{fig:5}(a) for $8$
sites. It is an $S=0$ state with momentum $\vec{k}=\vec{0}$ and is symmetric
with respect to spin inversion. The points of
maximal frustration are now around $0.57$ and $0.58$ respectively,
significantly changed from the classical and $s_{i}=7/2$ values. The
correlation functions are
plotted in Fig. \ref{fig:5}(b), and the change of the in-plane correlation
functions from a FM to a diagonal AF character is now smoother. This is
attributed to the smaller magnitude of the spins. The gain in
energy via the $J_{2}$ bond is now more significant.

\begin{figure}
\includegraphics[width=5in,height=5in]{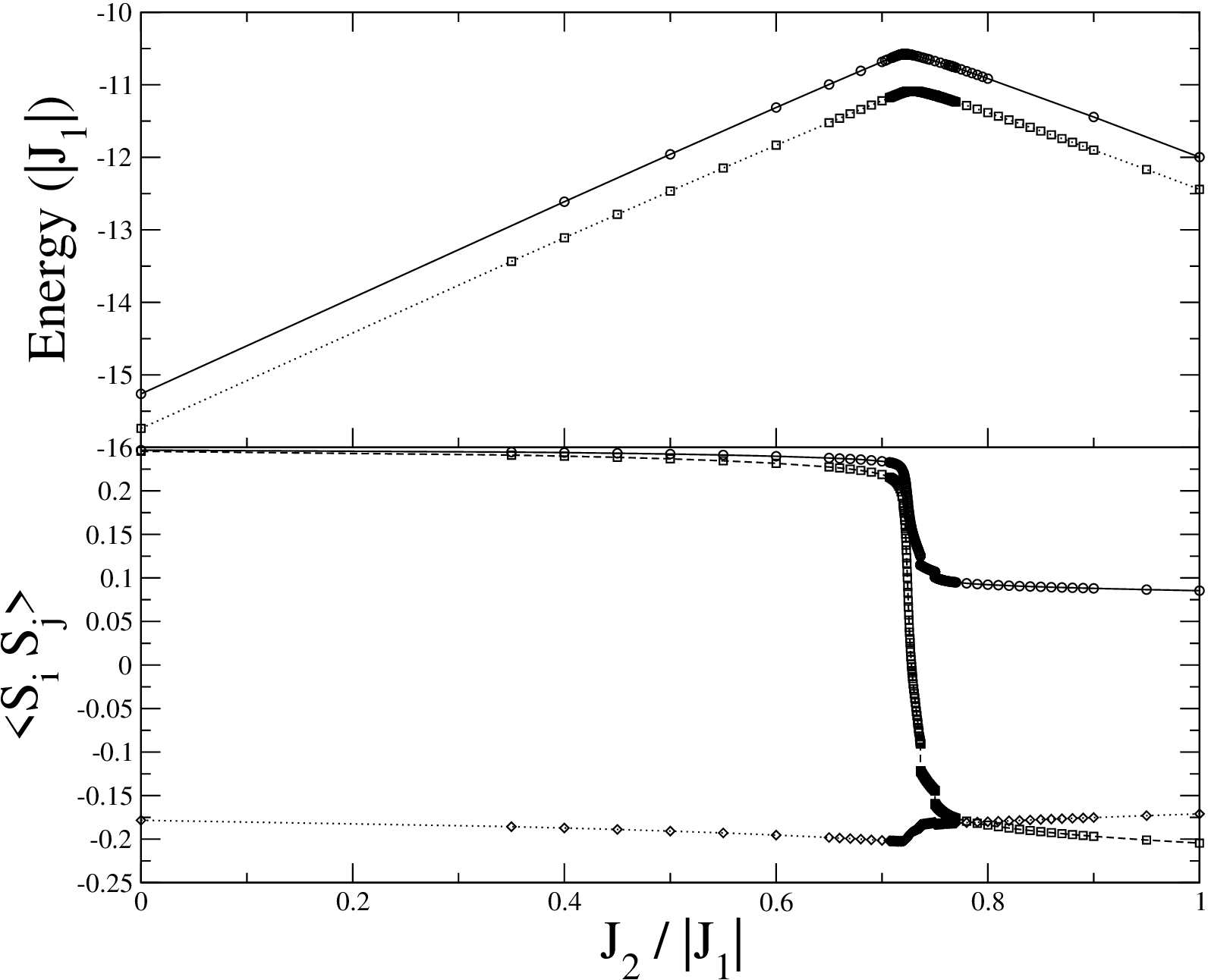}
\caption{Energy and nearest and next nearest neighbor correlation functions of
the ground state for $s_{i}=\frac{1}{2}$ and a $3 \times 3 \times 3$ cluster as
a function of $J_{2}/|J_{1}|$. Top: $J_{3}/|J_{1}|=0.4$ ($\circ$) and
$J_{3}/|J_{1}|=0.5$ ($\Box$). Bottom: correlation function for bond $J_{1}$
($\circ$), for bond $J_{2}$ ($\Box$), and for bond $J_{3}$ ($\diamond$). The
lines are guides for the eye.}
\label{fig:6}
\end{figure}

\begin{figure}
\includegraphics[width=5in,height=5in]{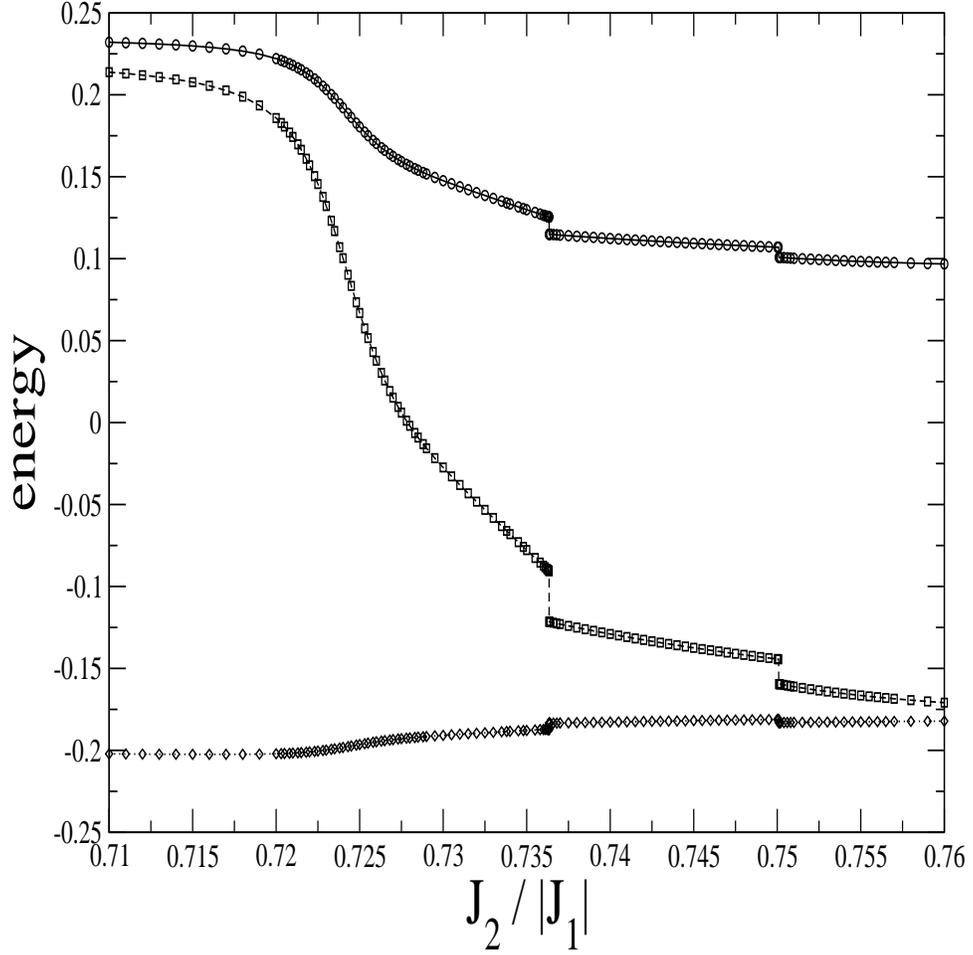}
\caption{Nearest and next nearest neighbor correlation functions of the ground
state for $s_{i}=\frac{1}{2}$ and a $3 \times 3 \times 3$ cluster as a function
of $J_{2}/|J_{1}|$ for $J_{3}/|J_{1}|=0.5$. $\circ$ : correlation function for
bond $J_{1}$, $\Box$: for bond $J_{2}$, $\diamond$ : for bond $J_{3}$. The
lines are guides for the eye.}
\label{fig:7}
\end{figure}

For the $27$-site cluster the ground state energy for $J_{3}/|J_{1}|=0.4$ and
$0.5$ is plotted in Fig. \ref{fig:6}(a) as a function of
$J_{2}/|J_{1}|$. Frustration is maximal for $J_{2}/|J_{1}|=0.7212$ and
$0.7278$. The ground state energy has momentum $\vec{k}=(0,0,\pi)$, which is
doubly degenerate. Its point symmetry group is $C_{2h}$, a subgroup of
$D_{2h}$. The ground state
belongs to the $A_{g}$ irreducible representation \cite{Altmann}. Correlation
functions for the ground state are plotted in Fig.\ref{fig:6}(b). Their
behavior is similar to the correlations in the $8$-site cluster and they change
character around the point of maximal frustration. The nearest neighbor
intraplanar correlation function is $0.247$
and almost fully polarized at $J_{2}/|J_{1}| = 0$, and then drops to approximately one third of this value for larger $J_{2}/|J_{1}|$. The diagonal correlation function starts with the same strong FM character at $J_{2}/|J_{1}| = 0$ and reverses sign for larger values. It is equal to $-0.205$ for $J_{2}/|J_{1}| = 1$, which shows strong AF correlation when compared with the value for $J_{1}=0$, which is $-0.238$. The interplanar correlation function originally decreases slightly with increasing $J_{2}/|J_{1}|$ and for higher values increases slightly to accommodate the increase of the diagonal intraplanar correlation function. Closer inspection of the plot (Fig. \ref{fig:7}) reveals two discontinuities in the correlation functions for values of $J_{2}/|J_{1}|$ equal to $0.736341$ and $0.75008$, where the two lowest energy states change roles as the ground and the first excited state. The two states have the same momentum and belong to the same irreducible representation, but the ground state between the two discontinuities is an $S=3/2$ state, while the other is $S=1/2$. At the same points the derivative of the energy with respect to $J_{2}/|J_{1}|$ is discontinuous.

As was the case for the $8$-site cluster, the transition from a FM to a spin
spiral state has been renormalized by quantum fluctuations, however the
corresponding $J_{2}/|J_{1}|$ value is closer to the classical value of $1$
compared with the $8$-site cluster. The changes of the intraplanar correlation
functions as functions of $J_{2}/|J_{1}|$ are sharper compared with the changes
in the $8$-site cluster. It is expected that the results for the $s_{i}=7/2$
case will be similar and the point where correlation functions change will be
closer to $1$, compared with the $8$-site case. This result agrees with linear
spin wave theory, where quantum fluctuations have a small effect on the
classical solution.

\begin{figure}
\includegraphics[width=5in,height=5in]{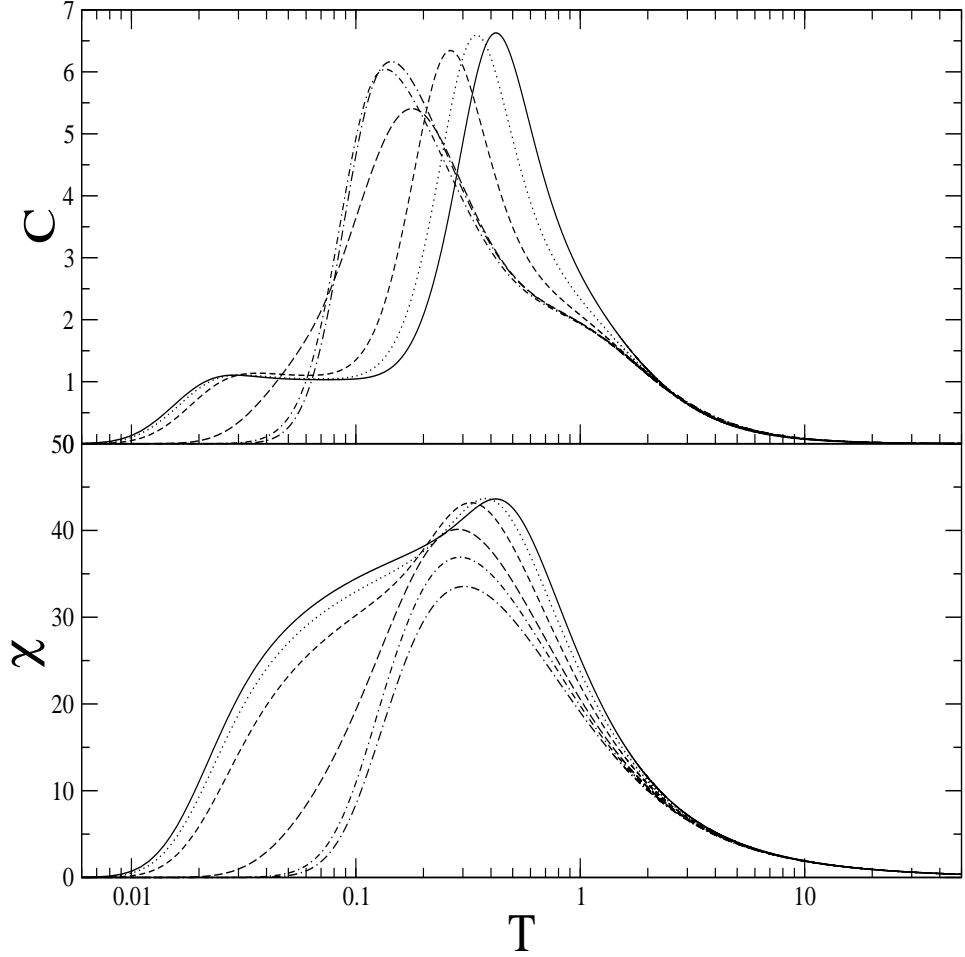}
\caption{Specific heat (top) and susceptibility (bottom) as function of
temperature for $J_{3}/|J_{1}|=0.5$. solid line : $J_{2}/|J_{1}|=0.4$, dotted
line : $J_{2}/|J_{1}|=0.5$, dashed line : $J_{2}/|J_{1}|=0.6$, long-dashed
line : $J_{2}/|J_{1}|=0.7$, dot-dashed line : $J_{2}/|J_{1}|=0.75$, dot-long
dashed line : $J_{2}/|J_{1}|=0.8$.}
\label{fig:8}
\end{figure}

When $s_{i}=1/2$ full diagonalization is possible for systems of 18 sites;
$3 \times 3 \times 2$ cluster Hamiltonians were diagonalized.
The specific heat of the cluster is plotted in Fig.
\ref{fig:8}(a) as a function of temperature for several values of $J_{2}/|J_{1}|$. There is a shoulder at low energy which disappears for $J_{2}/|J_{1}| > 0.7$, again indicating a change in the nature of the ground state. At the same time, the main peak is pushed towards lower temperature and its value decreases. The magnetic susceptibility as a function of temperature is shown in Fig. \ref{fig:8}(b). The shoulder found for lower $J_{2}/|J_{1}|$ values disappears for higher values and the peak position shifts to lower temperature, signifying again a qualitative change as the frustrating interaction gets stronger.

\section{Conclusions}
\label{sec:5}

A Heisenberg model with FM nearest neighbor interactions $J_{1}$ and AF
next nearest and interplanar interactions $J_{2}$ and $J_{3}$ has been studied as a prototype for
the magnetic behavior of polarons forming in the half-doped lanthanum manganite
\LaCa \cite{Zheng03}. The polarons are spin $7/2$ objects formed by two Mn and
an O ion. At the classical level the ground state is of the A-type for
$J_{2}/|J_{1}| \leq 0.5$, while for higher $J_{2}$ it is a spin spiral. When $J_{2}/|J_{1}| > 0.5$, there is a spiral in the $xy$ plane with every other pair of spins parallel in zig-zag chains, and an angle $q=arccos(-J_{1}/2J_{2})$ between neighboring spins. For $J_{2}/|J_{1}| \to \infty$ the spiral becomes the orthogonal state described by Efremov et al. in \cite{Efremov03}, where polaron moments are perpendicular to each other along the zig-zag chains and antiparallel along the direction of $J_{2}$. The effect of quantum fluctuations on spin wave theory is small due to the large magnitude of the spins, $s_{i}=7/2$. Hence the magnetization per site is not significantly changed from its classical value. Diagonalization of finite clusters also shows that quantum fluctuations do not significantly alter the classical solution, and the ratio $J_{2}/|J_{1}|$ for which there is a transition from FM order in the planes to one where spins are coupled antiferromagnetically via the $J_{2}$ bond is close to its value at the classical level.

The calculations in this paper show that the magnetic structure of Zener polarons in \LaCa is subtly dependent on the ratio $J_{2}/|J_{1}|$. The value from \ab calculations was found to be $0.38$ \cite{Zheng03}, however UHF calculations underestimate the value of AF couplings, and recent calculations of real space structure tend to favor a ratio close to $0.5$ with a CE-type magnetic ground state \cite{Patterson03}. Thus the physically relevant parameter space of the Hamiltonian (Eq. \ref{eqn:1}) has $J_{2}/|J_{1}| \sim 0.5$, where for $J_{2}/|J_{1}| > 0.5$ the magnetic structure is non-bipartite and depends on the exact value of $J_{2}/|J_{1}|$. Similar conclusions have been drawn by Efremov et al. in \cite{Efremov03}. In that paper the authors have found that around half-doping magnetic order depends sensitively on the extent of doping $x$, and the orthogonal state is the ground state for a part of the phase diagram, while a state which is a superposition of the conventional CE-type
 order and the orthogonal phase is lowest in energy for different combinations of parameters. The angle between neighboring spins is $2 \pi / 3$ in that phase. In our model the angle between spins linked via the $J_{2}$ bond and belonging to different zig-zag chains is $2 \pi / 3$ when $J_{2}/|J_{1}|=1$, while the angle between nearest neighbors is $\pi / 3$ in that case. This phase is also weakly renormalized by quantum fluctuations. Therefore the Hamiltonian (Eq. \ref{eqn:1}) predicts phases which differ from the conventional CE-type order.

Diagonalization of small clusters with cubic symmetry also showed that there is a change in the character of the ground state as a function of $J_{2}/|J_{1}|$. For small ratios spins are ferromagnetically ordered within planes, while for higher values the next-nearest neighbor AF interaction dominates. As was the case with linear spin wave theory, the diagonalizations show that the classical results are not significantly altered by quantum fluctuations. A spin magnitude of $s_{i}=1/2$ was also considered to study finite size effects, and the transition from the $8$- to the $27$-site cluster showed that the ratio where the character of the ground state changes comes closer to the classical value. Results from larger clusters would be needed to firmly establish this point, however memory requirements prohibit diagonalizations of larger cubic clusters.
The role of the AF interaction $J_{2}$ was also evident in
full diagonalizations of the model for systems of $18$ sites with $s_{i}=1/2$. Specific heat and magnetic susceptibility data were calculated, and the graphs change qualitatively as a function of $J_{2}/|J_{1}|$, showing again the role of the diagonal AF interaction in the development of AF intraplane correlations.

The spin wave dispersion was calculated for the material
Nd$_{0.45}$Sr$_{0.55}$MnO$_{3}$ in \cite{Yoshizawa98}, and it was fitted with a
Heisenberg model with FM nearest neighbor interactions in the planes
and AF interactions between planes. There was also an anisotropy
term. The relative strength of the AF with respect to the
FM interaction was found to be $0.620$, which is in agreement with
the values in \cite{Zheng03}. The anisotropic interaction was relatively small.

The authors wish to acknowledge discussions with R. Kenna. Calculations were
carried out at the Trinity Centre for High Performance Computing. NPK was
supported by a Marie Curie Fellowship of the European Community program
Development Host Fellowship under contract number HPMD-CT-2000-00048.

\bibliography{paperthree}

\end{document}